\documentclass{article}
\usepackage[utf8]{inputenc}
\usepackage{caption}
\usepackage{comment}
\usepackage{amssymb}
\usepackage{amsmath}
\usepackage{subfigure}
\usepackage{graphicx}
\usepackage{multirow}

\author{Kelly Boothby, Paul Bunyk, Jack Raymond, Aidan Roy}

\begin{document}

\title{Next-Generation Topology of D-Wave Quantum Processors}


\maketitle

\abstract{This paper presents an overview of the topology of D-Wave's next-generation quantum processors. It provides examples of minor embeddings and discusses performance of embedding algorithms for the new topology compared to the existing Chimera topology. It also presents some initial performance results for simple, standard Ising model classes of problems.}

\pagenumbering{arabic}

\section{Introduction}

This paper describes D-Wave's next-generation processor topology; that is, the
pattern that defines how the processor's qubits and couplers interconnect.
The flexible architecture of these new processors supports
simple design modifications able to produce various topologies: the \emph{Pegasus}
family of topologies.

Pegasus is a significant advancement over D-Wave's \emph{Chimera}
topology, which is available in the 2000Q product and its predecessors. Pegasus features
qubits of degree 15 and native $K_4$ and $K_{6,6}$ subgraphs.

Advantages of this new topology include:
\begin{itemize}
    \item more efficient embeddings of cliques, bicliques, 3D lattices and penalty models,
     and improved heuristic embedding run times;
    \item novel class of couplers that help error correction schemes, boosting energy scales and providing
     parity/auxiliary qubits; these are also useful in encoding various logical constraints.
\end{itemize}

\section{Pegasus Family of Topologies}\label{sec:family}

This section presents a general definition of processor topologies for the Pegasus family, \emph{Pegasus(x)}.
Where the document simply refers to ``Pegasus'', it is our target initial release, Pegasus(0).
Similar to our use of the $C_n$ notation for a Chimera graph with size parameter $n$, we refer to instances of Pegasus topologies by $P_n$; for example, $P_3$ is a
graph with 144 nodes.

\begin{figure}
 \centering
 \includegraphics{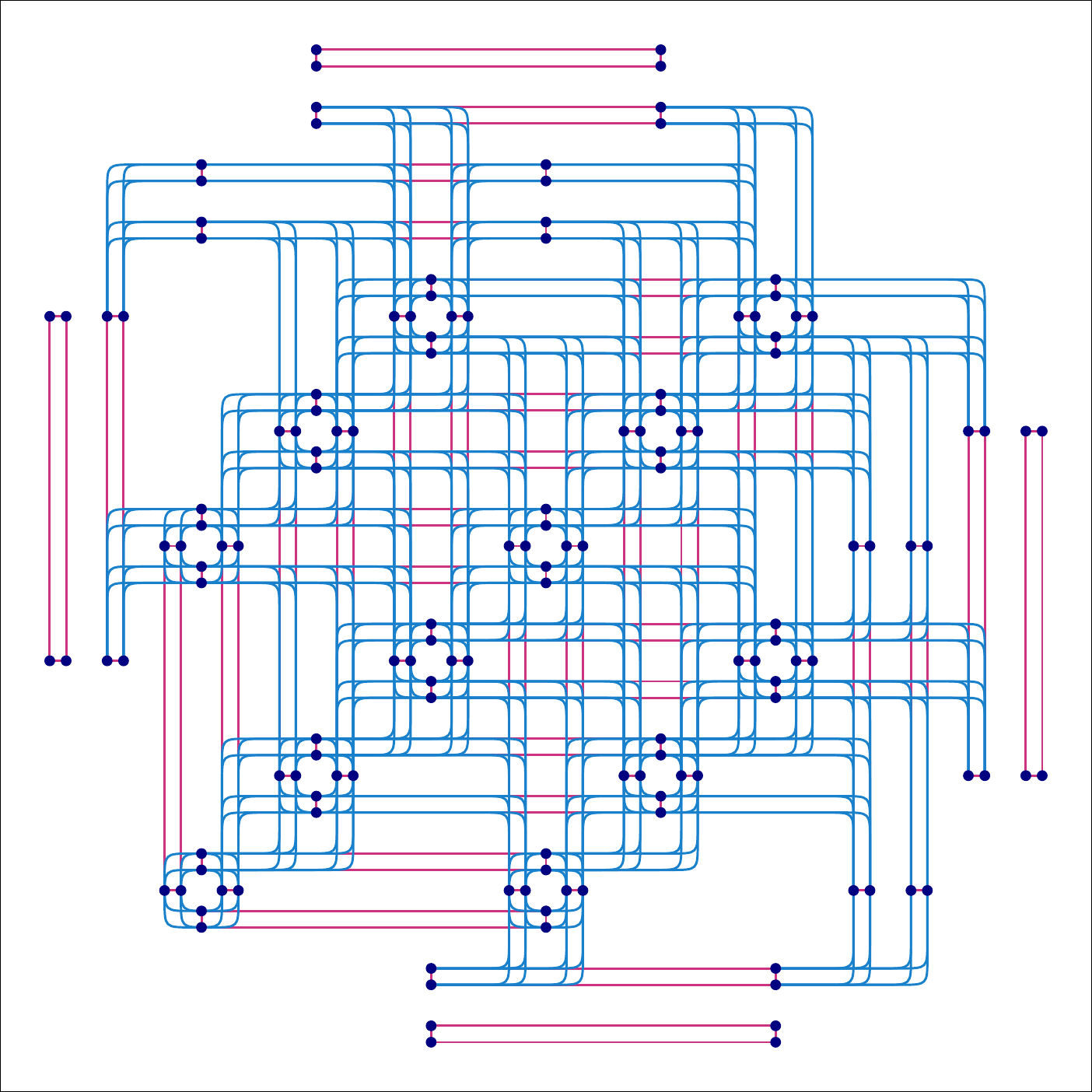}
 \caption{Roadway-style drawing of qubits (dots) and couplers (lines) in a $P_3$-sized Pegasus(0) processor, where
  curved blue lines are “internal” couplers, long red lines are “external” couplers, and short red lines are
  “odd” couplers.}
 \label{fig:construction}
\end{figure}

In Pegasus as in Chimera, qubits are ``oriented'' vertically or horizontally. In Chimera, there are two
types of coupler: internal couplers connect pairs of orthogonal (with
opposite orientation) qubits, and external couplers connect colinear pairs of qubits (that is,
pairs of qubits that are parallel, in the same row or column). The Pegasus family has, in addition
to Chimera's internal and external couplers, a third type: odd couplers. Odd couplers connect parallel
qubit pairs in adjacent rows or columns; see Figure~\ref{fig:construction}.

In the Chimera topology, qubits are considered to have a nominal length of 4 (each qubit
is connected to 4 orthogonal qubits through internal couplers) and degree of 6 (each qubit
is connected to 6 different qubits through couplers). In the Pegasus family, qubits have
a nominal length of 12 and degree of 15.

\begin{figure}
 \centering
 \includegraphics{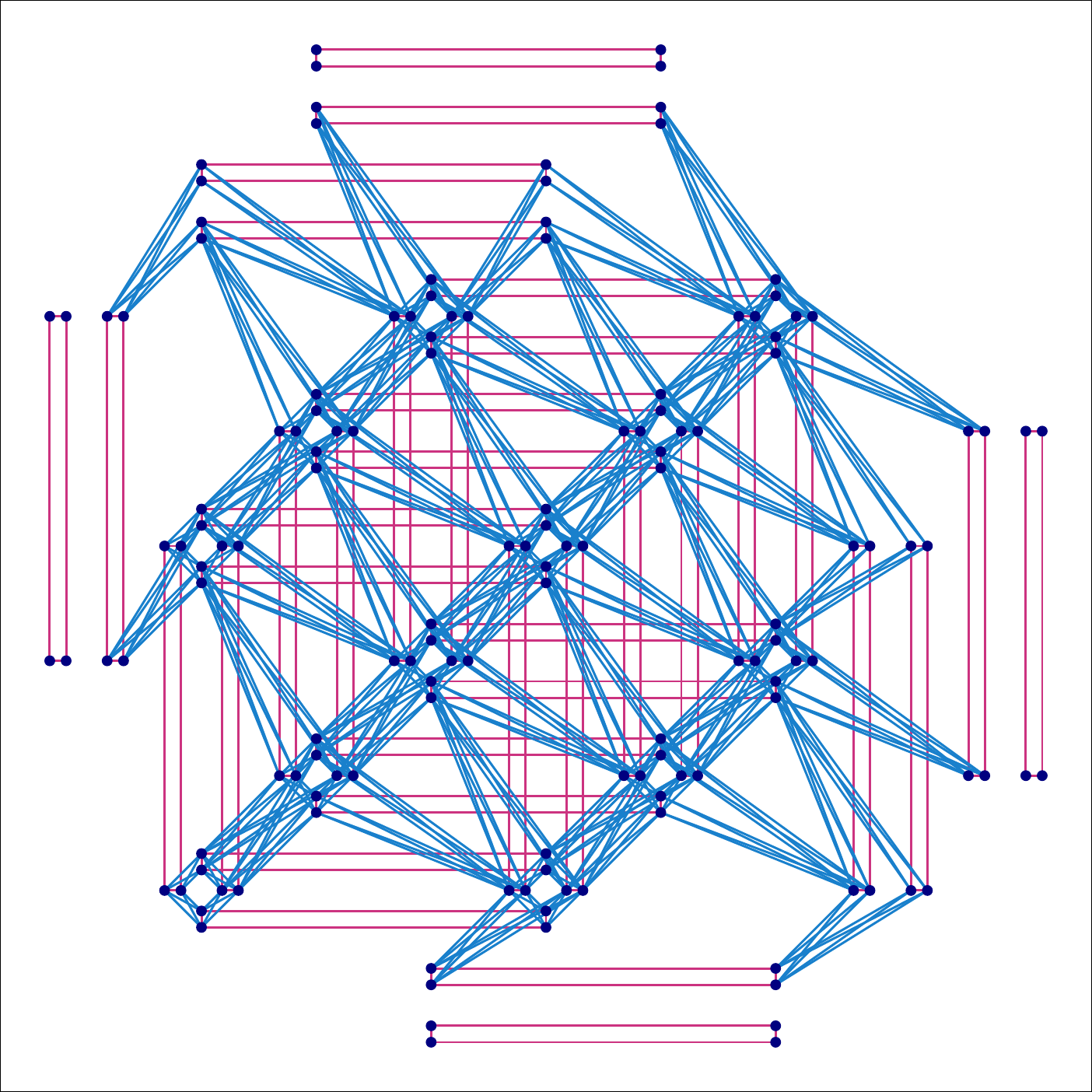}
 \caption{Straight-line drawing of qubits (dots) and couplers (lines) in a $P_3$-sized Pegasus(0) processor, where
  blue lines are “internal” couplers, long red lines are “external” couplers, and short red lines are
  “odd” couplers.}
 \label{fig:construction_straight}
\end{figure}

\subsection{Formulaic Description of Pegasus Topologies}

In broad strokes, a $P_M$ contains $24M(M-1)$ qubits, and has maximum degree 15.  The Pegasus(0) topology contains $8(M-1)$ qubits which are disconnected from the main processor fabric, for a total of $8(3M-1)(M-1)$ qubits in the main fabric.  For example, a $P_{16}$ contains 5760 qubits in total, and 5640 in the main fabric.

Let $M$ be a positive integer, and $s$ a vector of length 12 consisting of even values between $0$ and $10$ inclusive.  The contents of $s$ are called \emph{shifts}, and we write
$$s = (s^{(v)}_0, s^{(v)}_1, \cdots, s^{(v)}_5, s^{(h)}_0, s^{(h)}_1, \cdots, s^{(h)}_5),$$
separating $s$ into vertical $(v)$ and horizontal $(h)$ shifts.

The qubits for $P = P_M^{s}$ are defined by the cartesian product
\[
 V(P) = \{0,1\} \times \{0, \cdots, M-1\} \times \{0, \cdots, 11\} \times \{0, \cdots, M-2\}.
\]

\begin{figure}
 \centering
 \hspace*{-0.5cm}\includegraphics{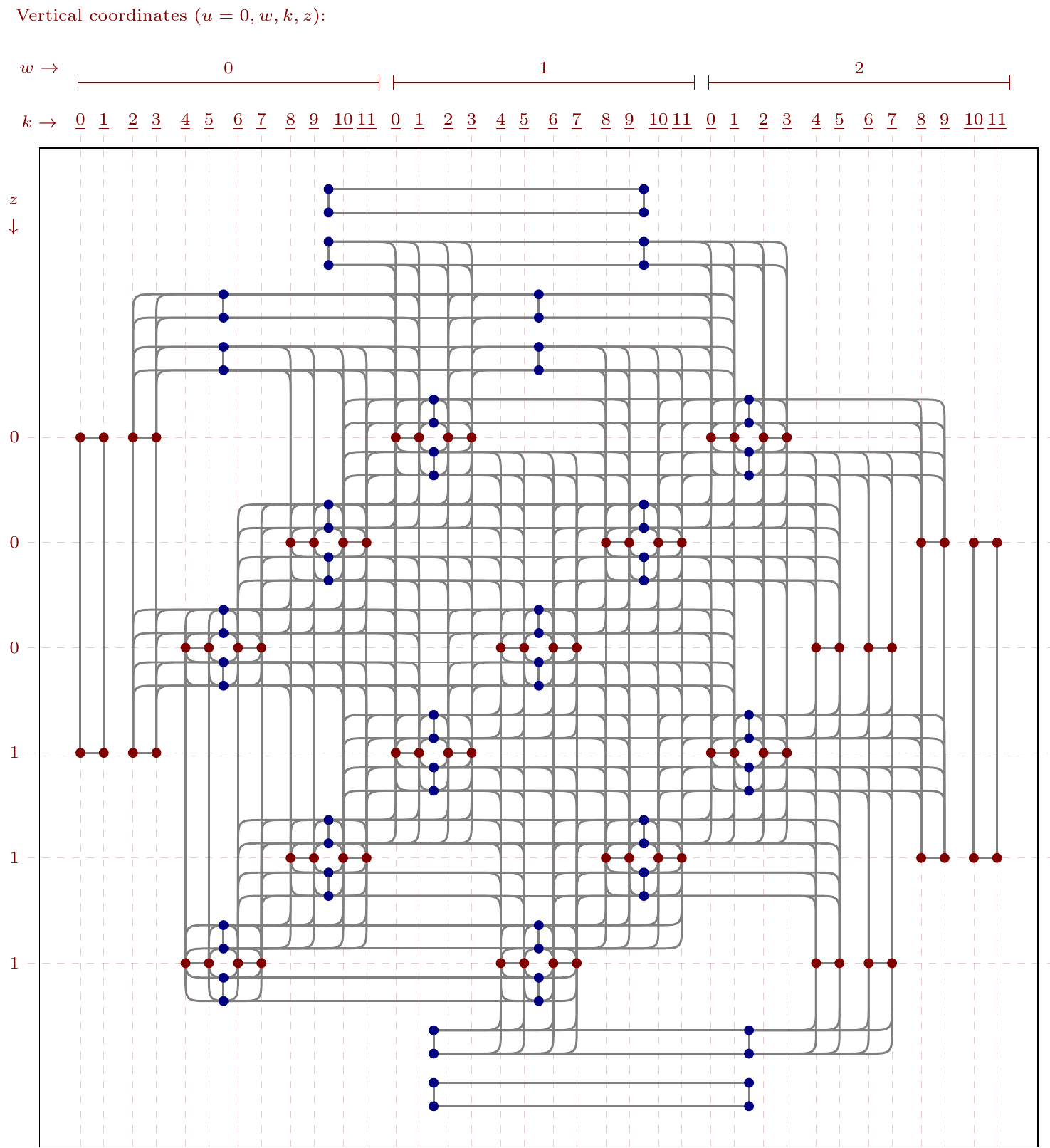}
 \caption{Coordinates of vertical qubits in a Pegasus $P_3$ processor.}
 \label{fig:coordinates_v}
\end{figure}

\begin{figure}
 \centering
 \hspace*{-1cm}\includegraphics{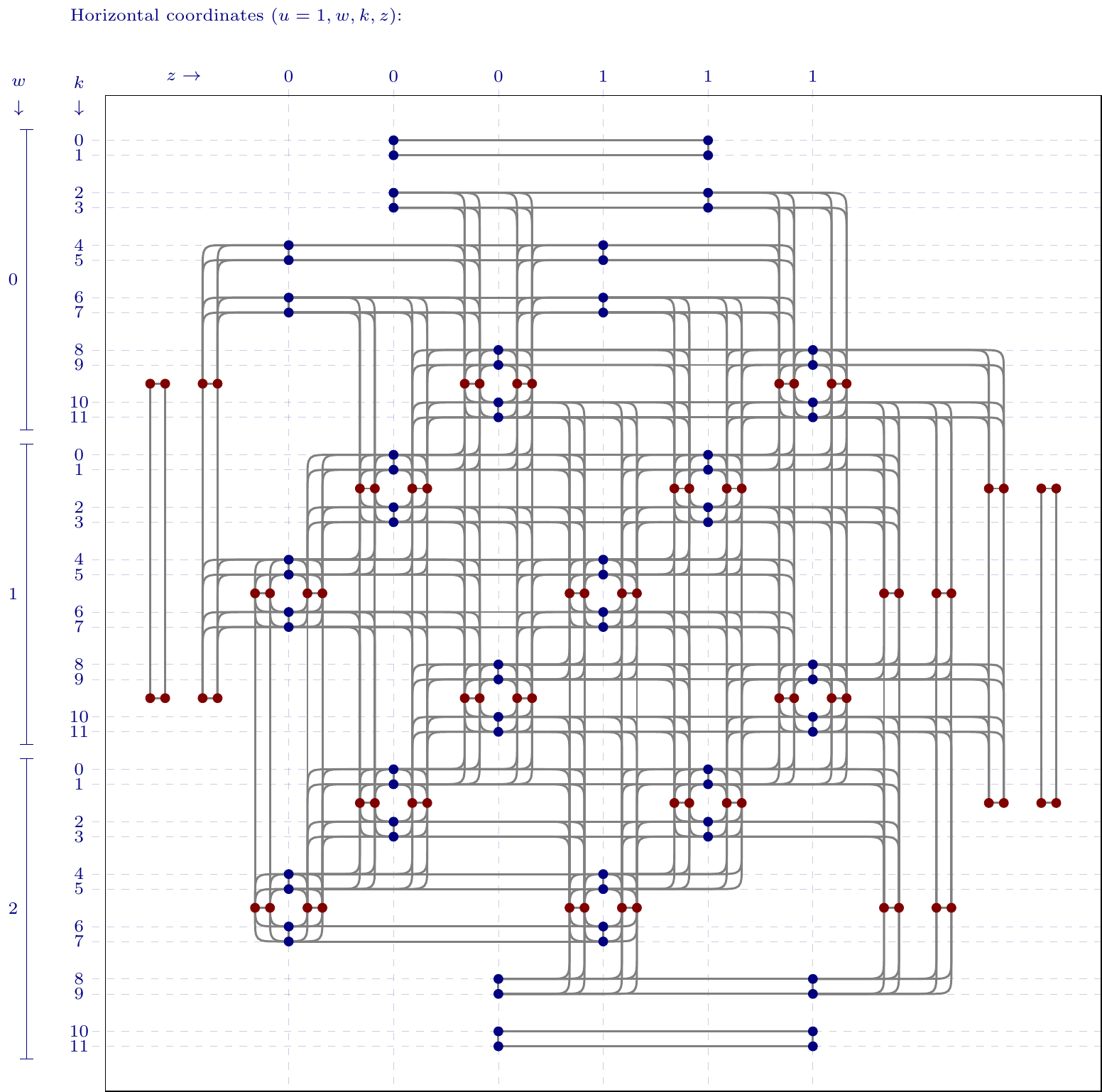}
 \caption{Coordinates of horizontal qubits in a Pegasus $P_3$ processor.}
 \label{fig:coordinates_h}
\end{figure}

For easier description, we name the coordinates of qubits. For a qubit $(u,w,k,z)$ in $V(P)$:
\begin{itemize}
\item $u$ is the \emph{orientation}, indicating if a qubit is vertical ($u=0$) or horizontal ($u=1$).
\item $w$ is the \emph{perpendicular tile offset}, indicating the index of the qubit's tile, in the orientation perpendicular to $u$. (That is, if $u=0$, then $w$ is a horizontal (column) index, and if $u=1$, then $w$ is a vertical (row) index.)
\item $k$ is the \emph{qubit offset}, indicating the index of a qubit within a tile.
\item $z$ is a the \emph{parallel tile offset}, indicating the index of the qubit's tile in the orientation parallel to $u$. (That is, if $u=0$, then $z$ is a vertical (row) index, and if $u=1$, then $z$ is a horizontal (column) index.)
\end{itemize}
These coordinates are shown in Figures \ref{fig:coordinates_v} and \ref{fig:coordinates_h}.

In the following description of the three types of couplers, a coupler $p \sim q$ exists whenever both $p$ and $q$ are contained in $V(P)$. Descriptions of the first two types are simple; the third includes a delta function, $\delta(a<b)$, defined as being equal to 1 if $a < b$ and to 0 otherwise. The three sets of Pegasus couplers are:

\begin{itemize}
 \item \emph{external}: $(u, w, k, z) \sim (u, w, k, z+1)$
 \item \emph{odd}: $(u, w, 2j, z) \sim (u, w, 2j+1, z)$
 \item \emph{internal}: $(0, w, k, z) \sim (1, z+\delta(j < s^{(v)}_{\lfloor k / 2 \rfloor}), j, w-\delta(k < s^{(h)}_{\lfloor j / 2\rfloor} ) )$
\end{itemize}

The Pegasus(0) topology is defined by $$s_0 = (2,2,10,10,6,6,6,6,2,2,10,10)$$ and written $P_M = P^{s_0}_M$.

A \emph{column} in Pegasus is indexed by a triple $(u=0,w,k)$ and a \emph{row} is indexed by a triple $(u=1,w,k)$. The qubits of a single row or column, namely $\{(u,w,k,z): 0 \leq z \leq M-2 \}$, form a path connected by external couplers, which is useful for creating chains.

We define an integer labeling for Pegasus with function \[(u, w, k, z) \mapsto z + (M-1)(k + 12(w + Mu)),\] which is a bijection between the $24M(M-1)$ coordinate labels and the interval \[\{0, \cdots, 24M(M-1)-1\}.\]

The main fabric of the processor, the largest connected component, occurs for those $(u, w, k, z)$ with
\[
u = 0 \qquad \text{and} \qquad \min_{0 \leq t < 6} s^{(h)}_t \leq 12w+k < 12(M-1)+\max_{0 \leq t < 6} s^{(h)}_t,
\]
or
\[
u = 1 \qquad \text{and} \qquad \min_{0 \leq t < 6} s^{(v)}_t \leq 12w+k < 12(M-1)+\max_{0 \leq t < 6} s^{(v)}_t.
\]

\section{Minor-Embedding}

Much of the embedding support that exists for the Chimera topology may be extended
to the Pegasus family of topologies with relative ease. Known constructions for Chimera embeddings
of structured problems translate to Pegasus without modification, because Chimera occurs as a subgraph of Pegasus.

For both structured and unstructured problems, heuristic, architecture-naive embedders consistently
produce shorter chains on Pegasus than on Chimera. Because longer chains increase errors in problem specification and reduce logical-variable fidelity, we expect that this
reduction in chain length will significantly improve problem solving.

\subsection{Clique and Biclique Embeddings}\label{sec:embclique}

\begin{figure}
  \centering
  \includegraphics{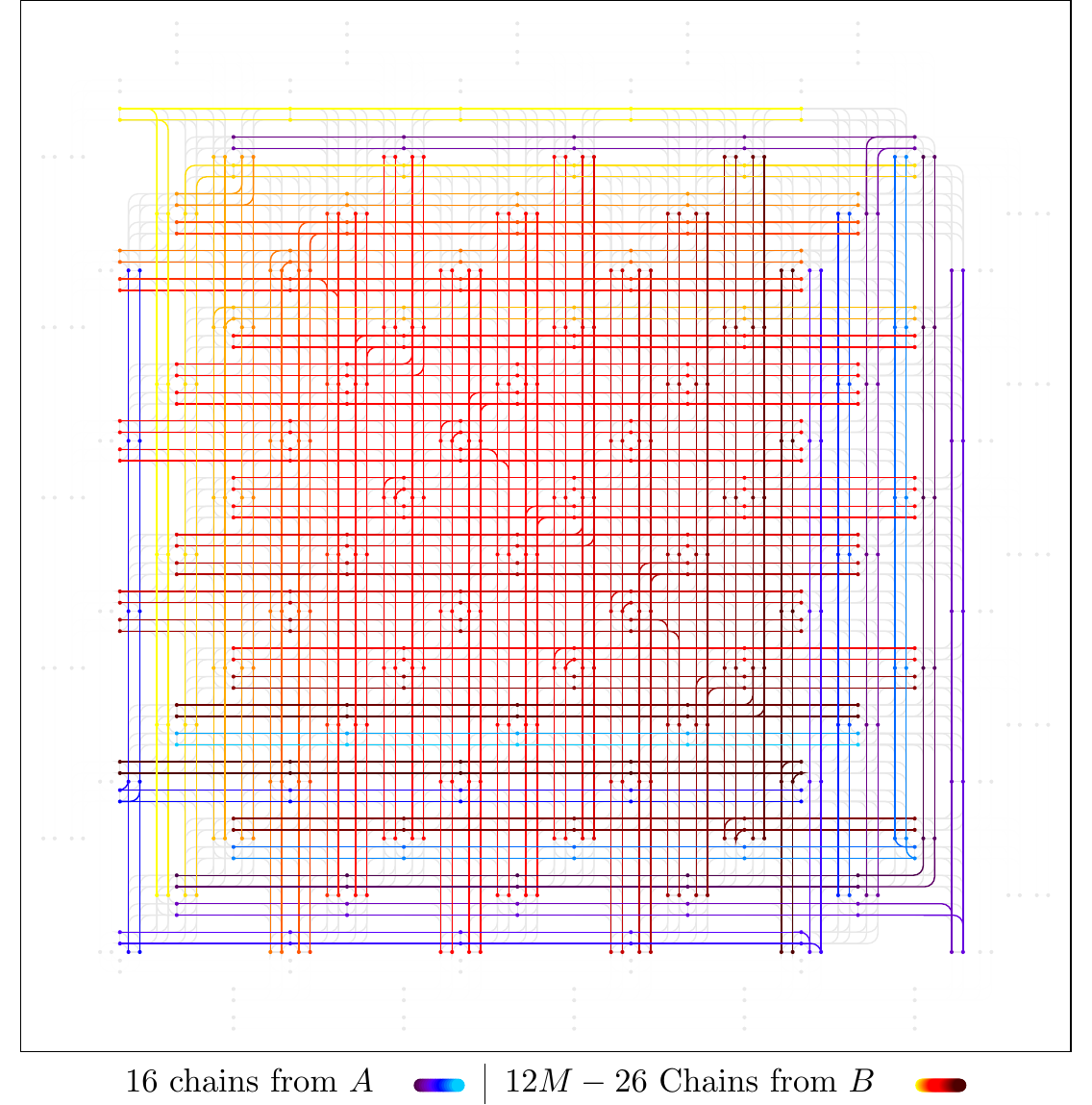}
  \caption{Embedding of $K_{62}$ in $P_6$.  See Section~\ref{sec:embclique}}\label{fig:clique}
\end{figure}

Embeddings of \emph{cliques} (complete graphs) and \emph{bicliques} (complete bipartite graphs) are very similar to those in Chimera.
This subsection briefly describes how to find those embeddings and reports maximum-yield formulas. We begin by finding elements common to both topologies, namely rows and columns of qubits, and describing how to assemble these into embeddings.

Optimal embeddings of complete bipartite graphs, $\varepsilon : K_{a,b} \to P$, consist of $a$ parallel paths of horizontal qubits, each in a different row, and $b$ parallel paths of vertical qubits, each in a different column. The maximum-sized embedding produced this way is $K_{12M-20, 12M-20}$ with uniform chain length of $M-1$.

Optimal embeddings of complete graphs, $\varepsilon : K_a \to P$ consist of $a$ parallel paths of horizontal qubits and $a$ parallel paths of vertical qubits, connected at a point of intersection to make $a$ chains. The algorithm of \cite{cliquepaper} is easily modified to produce embeddings with chains of length $M$ and $M+1$. Embeddings produced this way have size at most $a=12(M-1)$.

Allowing longer chains, the embedding strategy of \cite{KlySulHum} can produce cliques of size up to $a=12M-10$.  We describe such an embedding by making a set of \emph{chain descriptors} and then showing how to translate these into chains for an embedding. A chain descriptor is a set of triples $(u, w, j)$, each of which is expanded into one of two lines of qubits,
\[
 \{(u,w,2j,z) : 0 \leq z < M-1\} \text{ and } \{(u,w,2j+1,z) : 0 \leq z < M-1\}.
\]

For $M > 2$, let
\begin{align*}
      & A = \left\{
\{ (0, 0, 3), (1, M-2, 3) \},
\{ (0, M-2, 0), (1, M-2, 1) \},
\{ (0, M-2, 2) \},
 \right. \\
      & \hspace{1cm}
\left.
\{ (1, M-1, 4), (0, M-2, 3) \},
\{ (0, M-1, 4), (1, M-1, 5) \},
 \right. \\
      & \hspace{1cm}
\left.
\{ (1, 0, 4), (0, M-2, 1) \},
\{ (0, M-1, 5), (1, M-2, 0) \},
\{ (1, M-2, 5) \}
 \right\},
\end{align*}
giving a set of eight chain descriptors.  For the remaining $6M-13$ descriptors, define
$\rho = (4, 0, 2, 1, 3, 5)$, and
\[
 B = \left\{ \{(0, w, k), (1, w+\delta(k>1), \rho(k))\} :
 4 \leq 6w+k < 6M-9 \right\}.
\]
Our entire set of chain descriptors is given by $C = A \cup B$. To turn $C$ into an embedding, expand the descriptors into rows or columns of qubits,
\[
 E = \left\{ \{ (u, w, 2k+t, z) : (u,w,k) \in c, 0 \leq z < M-1 \} : c \in C, 0 \leq t < 2 \right\}.
\]
It is relatively straightforward to verify that the sets of qubits of $E$ form the chains of an embedding of $K_{12M-10}$. Figure~\ref{fig:clique} shows an illustration for $M=6$.

\subsection{Cubic Lattice Embedding}\label{sec:cube}

\begin{figure}
  \centering
  \includegraphics{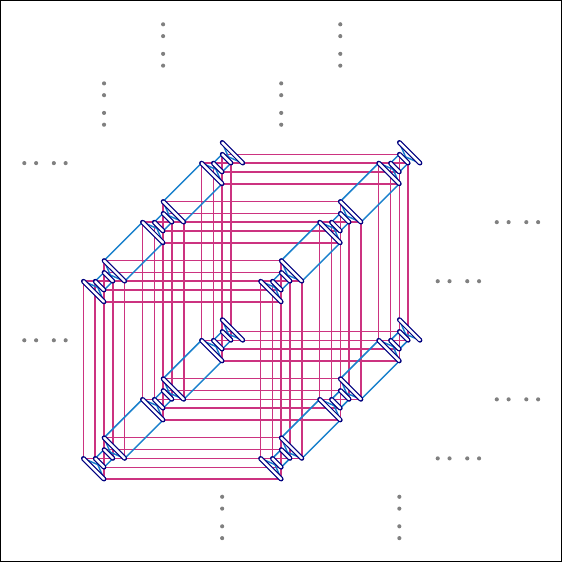}
  \caption{3d lattice embedding, $2 \times 2 \times 12 \to P_3$.  Chains are shown as elongated blue loops, and couplers associated with grid edges are shown as straight lines.}\label{fig:cube}
\end{figure}

As with clique and biclique embeddings, embedding the 3d cubic lattice is relatively straightforward with intuition from Chimera.  Figure~\ref{fig:cube} shows an embedding of a $2 \times 2 \times 12$ cubic lattice in $P_3$ with uniform chainlength of $2$; in larger instantiations of Pegasus, $P_M$, this embedding can grow to $(M-1) \times (M-1) \times 12$.
It is the simplest of a large number of cubic lattice embeddings,
\begin{itemize}
 \item $(x,y,z) \to \{(0,x,z+4,y),(1,y+1,7-z,x)\}$ for $0 \leq z < 8$, and
 \item $(x,y,z) \to \{(0,x+1,z-8,y),(1,y,19-z,x)\}$ for $8 \leq z < 12.$
\end{itemize}

\subsection{Selected 2d Lattice Embeddings}\label{sec:2dlat}

\begin{figure}[h]
  \centering
  \subfigure[Graphene lattice; a subgraph of $P^{s_0}_6$]{%
     \label{fig:hex}%
     \includegraphics[scale=1.2]{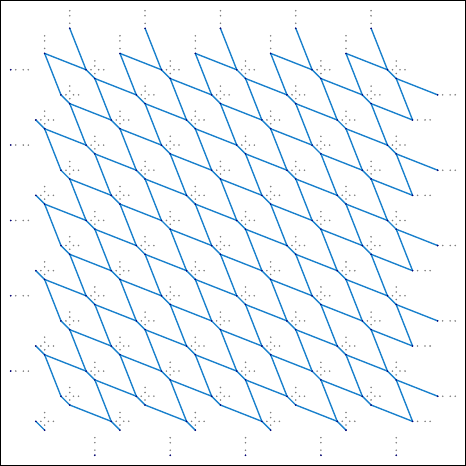}}%
     \qquad
  \subfigure[Square grid with diagonals, $\Gamma_{15}$ (embedding not shown, see Section~\ref{sec:2dlat})]{%
     \label{fig:mesh}%
     \includegraphics[scale=1.2]{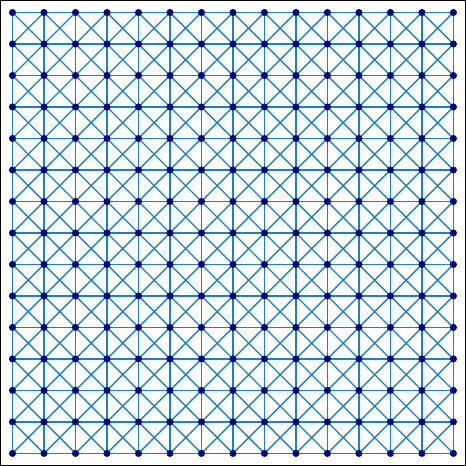}}%
  \caption{Two selected 2d lattices }
\end{figure}

Pegasus topologies enable many novel lattice embeddings. Embeddings like these appear to be useful for materials science; for example, computing magnetization phase diagrams.  Figure~\ref{fig:hex} shows a proper subgraph of Pegasus isomorphic to a finite portion of a graphene lattice.  A subgraph, of course, may be interpreted as an embedding with chainlength 1.  We obtain the subgraph $\Lambda$ shown in Figure~\ref{fig:hex} by deleting external couplers and taking the induced graph on set
\[ \{(u,w,k,z) \in V(P) : (u=1 \text{ and } k \in \{3,7,11\}) \text{ or } (u=0 \text{ and } k \in \{0,4,8\}) \}. \]

Taking a larger set,
\begin{eqnarray*}
 \{(u,w,k,z) \in V(P) & : & (u=1 \text{ and } k \in \{2,3,6,7,10,11\}) \text{ or } \\
                      &   & (u=0 \text{ and } k \in \{0,1,4,5,8,9\}) \},
\end{eqnarray*}
we obtain AA-stacked bilayer graphene, $K_2 \boxtimes \Lambda$ (where $\boxtimes$ denotes the strong graph product). Similarly,
\begin{eqnarray*}
 \{(u,w,k,z) \in V(P) & : & (u=1 \text{ and } k \in \{2,3,6,7,10,11\}) \text{ or } \\
                      &   & (u=0 \text{ and } k \in \{0,3,4,7,8,11\}) \}
\end{eqnarray*}
produces AB-stacked bilayer graphene. In the Chimera topology, these lattices require chains of length at least 2.

With chains of length 2 in Pegasus, we can construct an embedding for a grid lattice with horizontal, vertical and diagonal connections. An $(n \times n)$ grid is denoted $\Gamma_n$.
In Table~\ref{tab:gridplus} is the embedding $\varepsilon_{x,y}$ of a $3 \times 3$ subgrid, particularly the nodes $[3x, 3x+2] \times [3y, 3y+2]$.  Observe that $\varepsilon_{0,0}$ is an embedding of $\Gamma_{3} \to P_2$.  More generally, we can construct an embedding of $\Gamma_{3(M-1)} \to P_M$ as a union of subgrid embeddings, $\varepsilon_{x,y}$ for $0 \leq x < M-1$ and $0 \leq y < M-1$.  In Chimera, this lattice requires chains of length 6.

\begin{table}[]
\begin{center}
\begin{tabular}{l|l|l|l|}
\hline
           &  Row $3x$               & Row $3x+1$               & Row $3x + 2$\\
\hline
Column $3y$   & $\scriptstyle (0,x,2,y),(1,y,7,x)$   & $\scriptstyle (0,x,8,y),(1,y,6,x)$    & $\scriptstyle (0,x,7,y),(1,y+1,4,x)$ \\
\hline
Column $3y+1$ & $\scriptstyle (0,x+1,0,y),(1,y,2,x)$ & $\scriptstyle (0,x,11,y),(1,y+1,0,x)$ & $\scriptstyle (0,x,6,y),(1,y+1,3,x)$ \\
\hline
Column $3y+2$ & $\scriptstyle (0,x+1,3,y),(1,y,8,x)$ & $\scriptstyle (0,x,10,y),(1,y,11,x)$  & $\scriptstyle (0,x+1,4,y),(1,y,10,x)$ \\
\hline
\end{tabular}
\caption{Embedding of the grid-with-diagonal lattice $\Gamma_{3n}$}\label{tab:gridplus}
\end{center}
\end{table}

\section{Heuristic Embedding Results}\label{sec:faceoff}

This section presents a small-scale study to investigate performance trends in heuristic embedding.
Overall, we find that chains produced for Pegasus are on the order of 40\% of the lengths produced for Chimera
and runtimes see a similar improvement. The heuristic embedding algorithm used for this study is \verb!minorminer! version 0.1.3, denoted $\mathbf{A}$ below.

\begin{figure}
 \centering
 \includegraphics[width=0.5\textwidth]{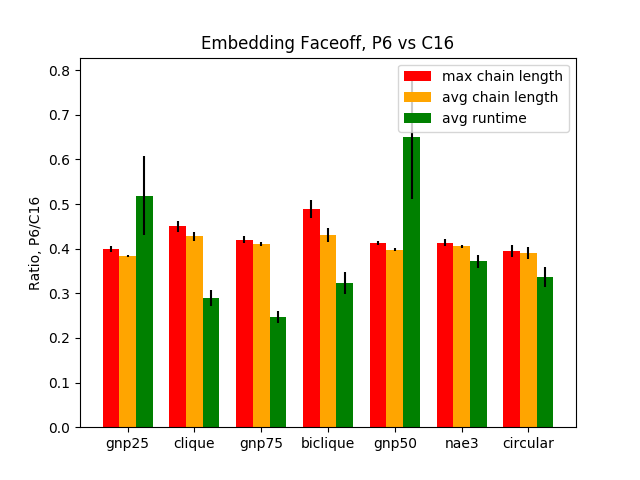}
 \caption{Embedding faceoff results, showing the mean and standard deviation of $F(m, S, P_6, C_{16})$ for $m = \ell, L, \tau$ and $S$ listed in Section~\ref{sec:problemsets}.}\label{fig:faceoff}
\end{figure}

\subsection{Methodology}
For problem $s$ and topology $T$, we write $c \in \mathbf{A}(s, T)$ for chain $c$ produced
by algorithm $\mathbf{A}$ for embedding $s \to T$.  For a fixed number of trials, $t$, we define three metrics,
\begin{itemize}
 \item \emph{average chainlength}: $\ell(s,T) = \frac{1}{t} \sum_{i=1}^t \frac{1}{|s|}\sum_{c \in \mathbf{A}(s,T)} |c|$,
 \item \emph{maximum chainlength}: $L(s,T) = \frac{1}{t} \sum_{i=1}^t \max_{c \in \mathbf{A}(s,T)} |c|$,
 \item \emph{average runtime}: $\tau(s,T)$ is the time taken to produce the $t$ embeddings, $\mathbf{A}(s,T)$, divided by $t$.
\end{itemize}
Note that heuristic embedding algorithms cannot generally be expected to produce embeddings every time.  To make these metrics sensible, we execute $\mathbf{A}(s,T)$ until we have accumulated $t$ embeddings and record the total time spent including failures.

Now, let $S$ be a set of problem graphs in a certain class.  An \emph{embedding faceoff} is a comparison between
two topologies, $T_1$ and $T_2$, for the three metrics above: $\mathbf{A}(s,T_1)$ versus $\mathbf{A}(s,T_2)$ for all $s \in S$.  Specifically, for a metric $m$, let
\[F(m, S, T_1, T_2) = \{ m(s, T_1)/m(s, T_2) : s \in S \}. \]

\subsection{Results}\label{sec:problemsets}

Figure~\ref{fig:faceoff} shows summary data for faceoffs between $P_6$ (with 680 qubits and 4484 couplers) and $C_{16}$ (with 2048 qubits and 6016 couplers) for each of the problem sets listed below, with $t=100$.  The following problem sets were chosen for their diversity, to show that these trends are not a result of properties of a particular problem set:

\begin{itemize}
 \item complete graphs, $K_n$ for $n=20$ to $n=29$ (labeled \verb!clique! in Figure~\ref{fig:faceoff}),
 \item complete bipartite graphs, $K_{n,n}$ for $n=22$ to $n=31$ (labeled \verb!biclique!),
 \item circular complete graphs, $K_{4n/n}$ for $n=10$ to $n=19$ --- these are graphs on $4n$ nodes $[0, 2n-1]$ with edges between $i$ and $i+n+j \mod 4n$ for $j=0, \cdots, 2n-1$, (labeled \verb!circular!),
 \item not-all-equal-3SAT graphs near the critical threshold; 10 instances with size 35 (labeled \verb!nae3sat!),
 \item Erd\"os-R\'enyi random graphs, $G(n,p)$, with 10 instances each of
 \begin{itemize}
  \item $G(70,.25)$ (labeled \verb!gnp25!),
  \item $G(60,.50)$ (labeled \verb!gnp50!), and
  \item $G(50,.75)$ (labeled \verb!gnp75!).
 \end{itemize}
\end{itemize}

Despite the $P_6$ having fewer qubits and couplers than the $C_{16}$, Pegasus consistently achieves around a 50-60\% reduction in chainlength over Chimera.

\section{Treewidth}

\begin{figure}
 \centering
 \includegraphics{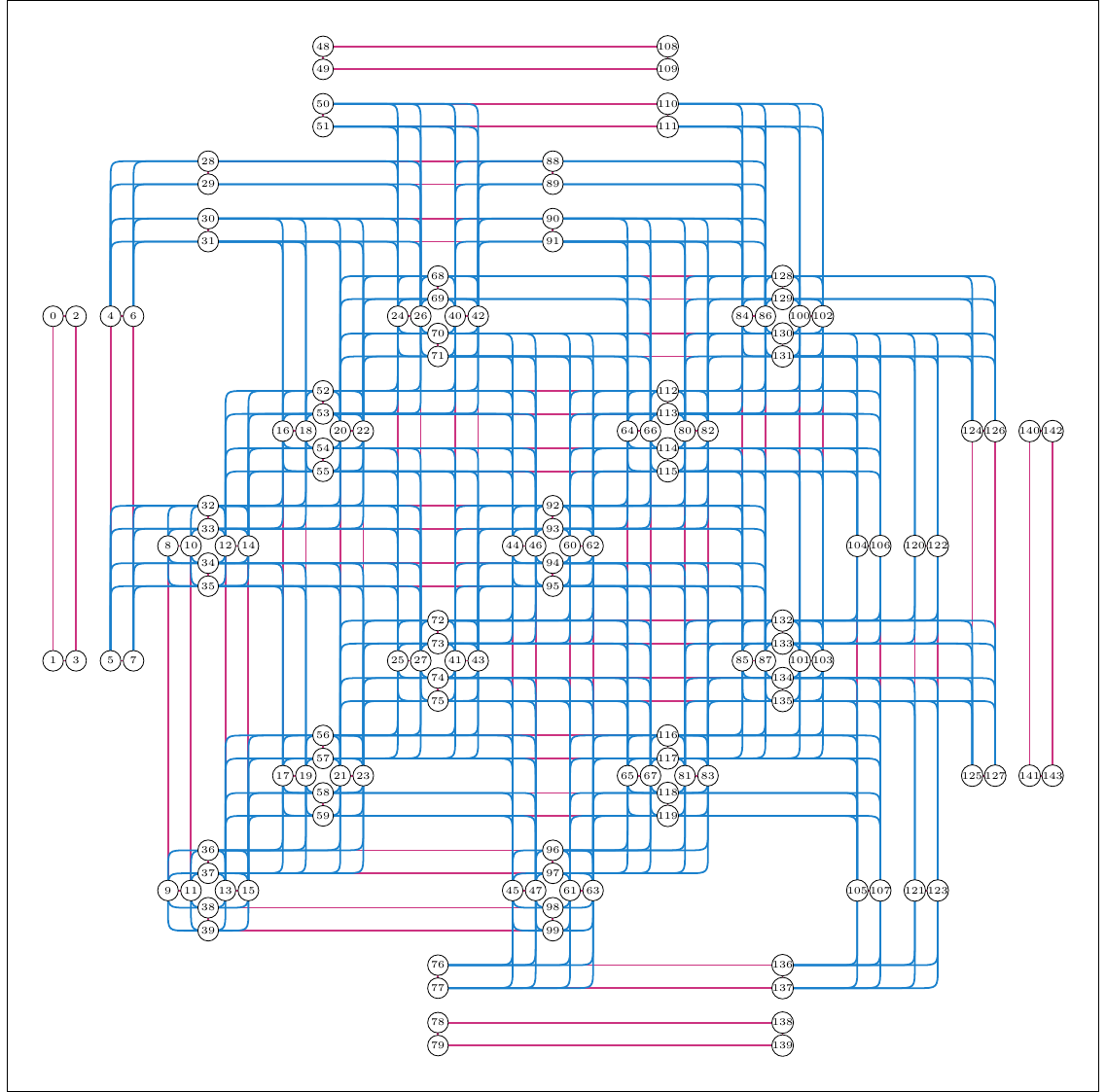}
 \caption{A vertex elimination order for $P_3$ with an elimination order width of $32$.}\label{fig:varorder}
\end{figure}

One measure of the complexity of a graph is its treewidth \cite{Robertson1986}. For example, the minimum energy of an Ising model defined on a graph of treewidth $t$ with $n$ vertices can be found in time $O(n2^t)$ using dynamic programming \cite{Dechter99bucketelimination}. Here we show that the treewidth of the Pegasus $P_M$ graph is between $12M-11$ and $12M-4$. For comparison, the treewidth of a Chimera $C_M$ graph is $4M$. In both cases, the treewidth is roughly the number of rows (or columns) of qubits as described in Section \ref{sec:embclique}.

To lower-bound the treewidth, consider embeddable complete graphs. A complete graph $K_n$ has treewidth $n-1$, and complete graphs of size $12M-10$ can be embedded in $P_M$. It follows that the treewidth of $P_M$ is at least $12M-11$ (see for example \cite[Lemma 15]{Bodlaender2011}). To upper-bound the treewidth, we provide a vertex elimination order with an elimination order width of $12M-4$ (see \cite[Theorem 6]{Bodlaender2010} for background). One such variable elimination order is as follows:
\begin{itemize}
\item eliminate vertical qubits, one parallel path column at a time;
\item eliminate horizontal qubits in any column once all vertical qubits adjacent to them have been eliminated.
\end{itemize}
An example of this variable elimination order for $P_3$ is given in Figure~\ref{fig:varorder}. It is straightforward to verify that the width of this order is $12M-4$.

\section{Error Correction}
\label{sec:error_correction}

The additional connectivity of the Pegasus graph can be used to construct a simple error correction scheme for quantum annealing. As in \cite{Pudenz2014} and \cite{Vinci2016}, we use multiple physical qubits to encode a single logical qubit in a way that increases the energy scale of the logical Ising problem.

\begin{figure}
  \centering
  \subfigure[]{%
      \includegraphics[width=0.7\textwidth]{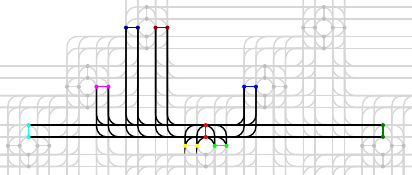}}%
  \\
  \subfigure[]{%
      \includegraphics[width=0.7\textwidth]{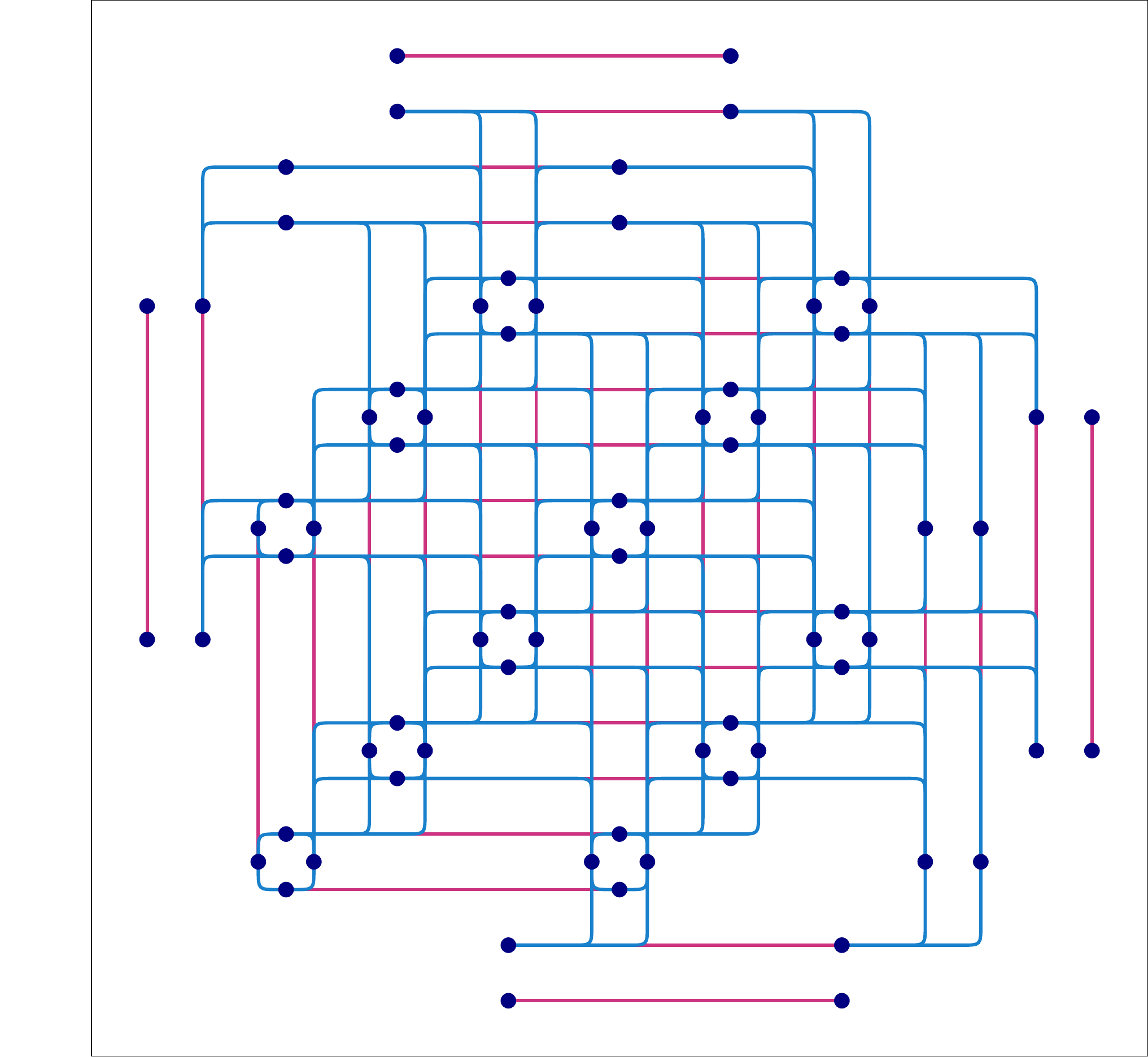}%
      \hspace{0.25in}} 
  \caption{A simple error correction scheme for Pegasus, in which pairs of qubits joined by an odd coupler are identified as a single logical qubit. In (a), each logical qubit is represented by a different color. In the resulting logical graph (b), internal couplers are represented four times physically, and external couplers are represented twice.}\label{fig:error_correction}
\end{figure}

In particular, let each pair of qubits joined by an odd-coupler represent a single logical qubit. The resulting logical graph is similar to Pegasus but with half as many qubits, no odd couplers, and typical degree $8$ (see Figure \ref{fig:error_correction}). From a classical error-correction perspective, this scheme is a simple repetition code: maintaining two copies of every variable allows for error detection (when the copies are not equal) as well as error suppression (chained qubits are likely to settle in the same state). However, because internal logical couplers are represented by four physical couplers, and external logical couplers are represented by two physical couplers, the energy scale at which the logical Ising model is represented is doubled.

If external couplers are used exclusively to create chains in logical graphs (as Section \ref{sec:embclique} embeddings demonstrate for the cliques and bicliques), then logical problem interactions are represented only using internal couplers, and this error correction scheme quadruples the energy scale of the problem. 

\section{Topology Implications for Native Structured Ising Models}

This section presents results of testing phase-transition properties on simple and well
understood benchmarks, as well as time-to-solution (TTS) results for standard classical algorithms, on Pegasus and Chimera topologies.\footnote{TTS is defined in this paper as the time required to solve a single problem instance with 50\% probability.}
Ultimately the performance of these topologies will be assessed in the context of hard optimization and inference problems. Regardless of what problems these prove to be, it is interesting as a first step to understand the interaction between the topology and the simplest standard Ising model classes it can express while exploiting all degrees of freedom (without embedding).

We also undertook simple studies of the impact of accelerator moves over these classes, namely Houdayer moves and large-area local-search moves. We find that, within these classes, the impact of accelerator moves is diminished in Pegasus---it appears to be more difficult to accelerate optimization in random Pegasus problems compared to random Chimera problems. TTS is also longer in Pegasus in typical cases across a range of methods.

Here we consider models defined by a problem Hamiltonian $H_P(x) = \sum_{ij} J_{ij} x_i x_j$. To minimize the impact of boundary effects, we use tori in all studies. The three cases discussed are:
\begin{itemize}
    \item Ferromagnet: $J_{ij}=-1,\; \forall\; ij$
    \item Maximum energy scale spin glass: $J_{ij} \in \{-1,1\},\; \forall\; ij$ (independent and identically distributed,    i.i.d.), which is called RAN1
    \item Maximum entropy spin glass: $J_{ij} \in [-1,1],\; \forall\; ij$ i.i.d., which is called RANinf
\end{itemize}

\subsection{Equilibrium Results in Ferromagnets and Spin Glasses}

Ferromagnetic phase transitions may superficially appear irrelevant to algorithmic performance in hard applications because ferromagnets are easy. Nevertheless, these transitions provide some intuition and bounds on the performance of local search methods. Models with larger transition values require stronger thermal or quantum excitations to explore the phase space by local search, due to larger (free) energy barriers. Since no problem orders more strongly than a ferromagnet, the ferromagnetic value provides a bound on the region in which we may see slow dynamics. The ferromagnetic transition can also be informative on the dynamics of large unfrustrated subdomains (or Griffiths singularities) within random problems, which can slow local-search methods in hard problems~\cite{PhysRevE.96.022139}.

Spin-glass transitions, where they exist, are expected to present a hard barrier to local-search methods, and an opportunity to demonstrate differentiation in dynamics.
The classical spin-glass phase transition is expected to be zero for RAN1 and RANinf in these graphs, and this has been argued to be one limitation in their use as benchmarks~\cite{PhysRevX.4.021008}.
However, we must bear in mind that even in the case of a zero-temperature transition, practical problems are finite, and spin-glass-like ordering can impact algorithms. We find at the large-system limit that spin-glass ordering is restricted to $T\rightarrow 0$, but this decay to zero is more rapid in Chimera than in Pegasus, so the landscape in Pegasus is in some sense rougher at a given low temperature.\footnote{This result is apparent in the scaling form, and will be presented in a separate document.}

Phase transitions are defined by singularities of the free energy $F(T,A) = -T\log \mathrm{-\frac{1}{T}{\hat H}}$ in the large-system limit, where ${\hat H}=\sum_{ij} J_{ij} \sigma^z_i \sigma^z_j - A \sum \sigma^x_i$. The classical transition is defined for $A=0$, whereas the quantum transition is defined in the limit $T\rightarrow 0$. Based on collapses of the Binder cumulant for the order parameter, we find (and plan to report elsewhere), the preliminary values for phase transitions shown in Table~\ref{tab:bindercollapses}. In this table, the maximum size is 1176 for Pegasus and 1152 for Chimera and the number of models ranges from 1 to 400.

\begin{table}[]
\begin{center}
\begin{tabular}{llll}
\hline
\multicolumn{1}{|l|}{Transition Type}  & \multicolumn{1}{l|}{On} & \multicolumn{1}{l|}{Models} & \multicolumn{1}{l|}{Critical Value} \\ \hline
\multicolumn{1}{|l|}{\multirow{2}{*}{Classical ferromagnet}} & \multicolumn{1}{l|}{Pegasus}  & \multicolumn{1}{l|}{1} & \multicolumn{1}{l|}{$T_c = 12.6$}  \\ \cline{2-4}
\multicolumn{1}{|l|}{} & \multicolumn{1}{l|}{Chimera}   & \multicolumn{1}{l|}{1}  & \multicolumn{1}{l|}{$T_C=4.16$~\cite{PhysRevX.4.021008,PhysRevX.5.019901}} \\ \hline
\multicolumn{1}{|l|}{\multirow{2}{*}{Quantum ferromagnet}}   & \multicolumn{1}{l|}{Pegasus}  & \multicolumn{1}{l|}{1} & \multicolumn{1}{l|}{$A_C=13.85$} \\ \cline{2-4}
\multicolumn{1}{|l|}{} & \multicolumn{1}{l|}{Chimera}   & \multicolumn{1}{l|}{1} & \multicolumn{1}{l|}{$A_C=5.05$} \\ \hline
\multicolumn{1}{|l|}{\multirow{2}{*}{Classical spin-glass}}  & \multicolumn{1}{l|}{Pegasus}  & \multicolumn{1}{l|}{400}     & \multicolumn{1}{l|}{$T_C = 0$}  \\ \cline{2-4}
\multicolumn{1}{|l|}{} & \multicolumn{1}{l|}{Chimera}  & \multicolumn{1}{l|}{400}  & \multicolumn{1}{l|}{$T_C = 0$~\cite{PhysRevX.4.021008}} \\ \hline
\end{tabular}
\caption{Phase transitions based on Binder-cumulant collapses for models with up to 1176-qubits in Pegasus and 1152-qubits in Chimera}\label{tab:bindercollapses}
\end{center}
\end{table}

Critical behavior is consistent with known exponents. At the time of writing we have not completed thorough studies on the quantum spin-glass transition, which is expected to be finite and larger in Pegasus than in Chimera, and scalings qualitatively similar to the 2d square lattice are anticipated~\cite{PhysRevLett.72.4141}.
Details of the methods will be presented elsewhere.

\subsection{Time-to-Solution Results in Spin Glasses}

This section considers several standard and state-of-the-art classical optimization methods for random disordered models:

\begin{itemize}
    \item Greedy descent: a simple algorithm that descends the energy landscape by single bit flips.
    \item Simulated annealing: a standard heuristic algorithm that exploits thermal excitations to search a classical energy landscape and relax upon a minimizing solution~\cite{Kirkpatrick:OSA}.
    \item Parallel tempering: an algorithm that explores the landscape in multiple parallel local searches, each with differing degrees of thermal excitations~\cite{doi:10.1143/JPSJ.65.1604}.
\end{itemize}

Performance of these methods on Chimera-structured spin-glasses is known to be strongly enhanced by incorporating large-area local-search moves~\cite{1409.3934} or Houdayer moves~\cite{PhysRevLett.115.077201}. This subsection demonstrates that the impact of these accelerator methods is significantly reduced for the Pegasus topology.
TTS results presented in the accompanying figures are restricted to typical case performance (median with respect to instance behavior), but qualitatively similar patterns are present in other quantiles.

Greedy descent quickly fails to determine the minima as system size increases. We consider instead a Hamze de-Freitas Selby (HFS) greedy descent method~\cite{Hamze2004,1409.3934}, using $N$ random trees of treewidth 14, determined by the minimum-degree heuristic~\cite{Gogate2004}. We perform an exact minimization over variables in a randomly selected tree, and iterate until energy no longer decreases---specifically until every variable has been optimized at least once since the last energy decrease. A related method has been found to be very successful on a variety of Chimera structured benchmarks~\cite{1409.3934}, the difference here being that we choose a standard heuristic for tree construction that is applicable to any graph, which less fully exploits the lattice symmetry.

\begin{figure}
  \centering
\includegraphics[width=0.33\linewidth]{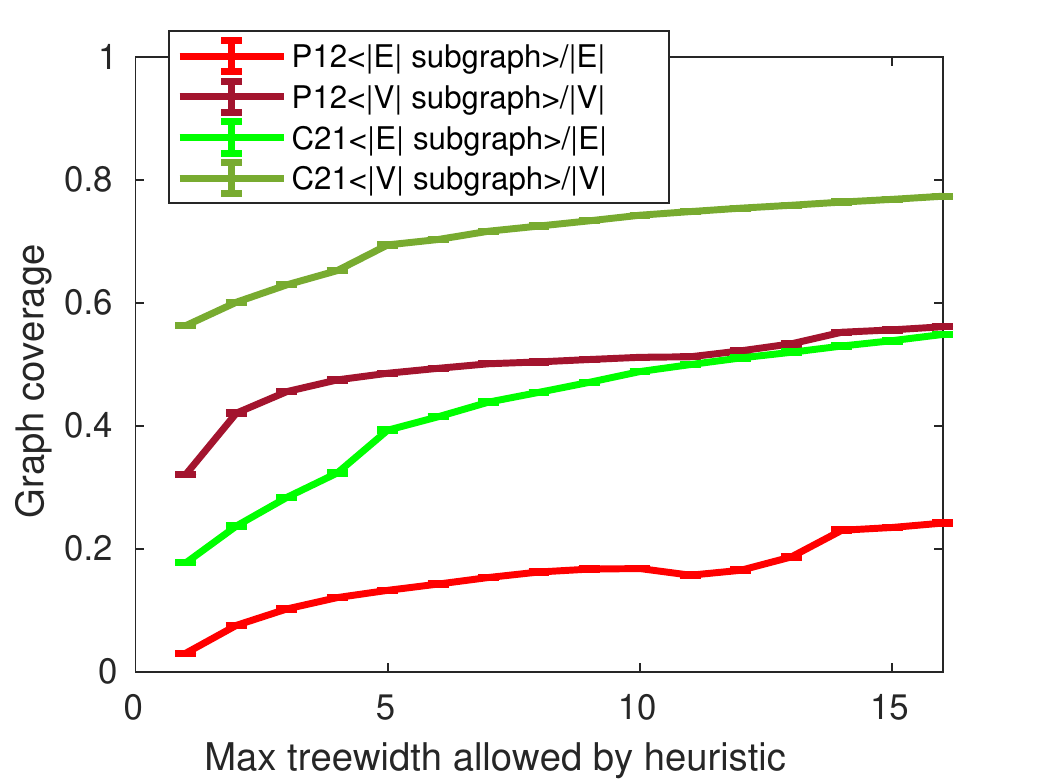}
\includegraphics[width=0.33\linewidth]{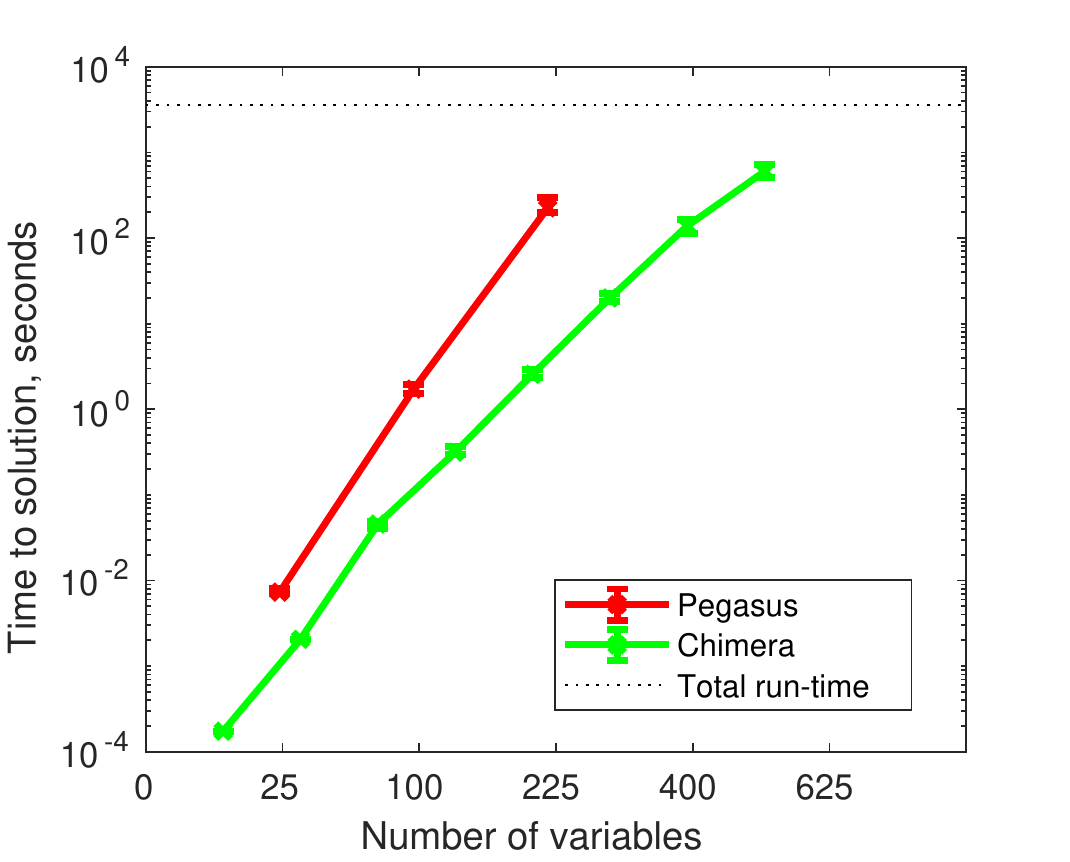}
\caption{\label{figureAlgs2} Iterations over graphs determined by the minimum-degree heuristic in HFS greedy descent.
  Left figure: Effectiveness of the minimum-degree heuristic, as measured by the mean fraction of vertices and edges that are contained entirely in the tree that is marginalized. We see a cusp for Chimera at treewidth 5, reflecting the ability of the algorithm to begin sampling over full cells, whereas a cusp for Pegasus occurs at treewidth 14.
  Right figure: Median time to solution for HFS greedy descent over 100 random instances of RANinf at each system size; TTS is significantly longer in Pegasus than in Chimera at comparable size. Treewidth was optimized (from 1 to 16) over a test set. Error bars are determined by a $90\%$ bootstrap-confidence interval with respect to instances. The implementation combined MATLAB with a C inner loop on single threaded 2.6 GHz CPUs, and is suitable for generic sparse graphs, although significant optimizations are possible for smaller treewidth graphs. System size is shown on a quadratic scale (proportional to the lattice linear dimension).}
\end{figure}

Figure \ref{figureAlgs2}, left side, shows the properties of the subgraphs used in the greedy descent, specifically the number of vertices $|V|$ and edges $|E|$ within the subgraphs compared to the full graph. Measuring the subgraphs produced by the minimum-degree heuristic method, we notice changes of behavior at treewidth 5 in Chimera and treewidth 14 in Pegasus. These are not sensitive to lattice size, except for in very small graphs, and reflect, in some sense, sweet spots for decomposition algorithms due to the cellular nature of the graphs. Figure \ref{figureAlgs2}, right side, shows the typical TTS where treewidth has been optimized. With this optimization in place, we find that the RANinf Pegasus benchmarks are significantly harder to optimize at equivalent size.

For simulated annealing we consider two variations on the update rule:

\begin{itemize}
    \item SA: Metropolis algorithm~\cite{Kirkpatrick:OSA}
    \item SA(CG): eight-qubit conditional Gibbs sampling, an accelerator
\end{itemize}

\begin{figure}
\includegraphics[width=0.33\linewidth]{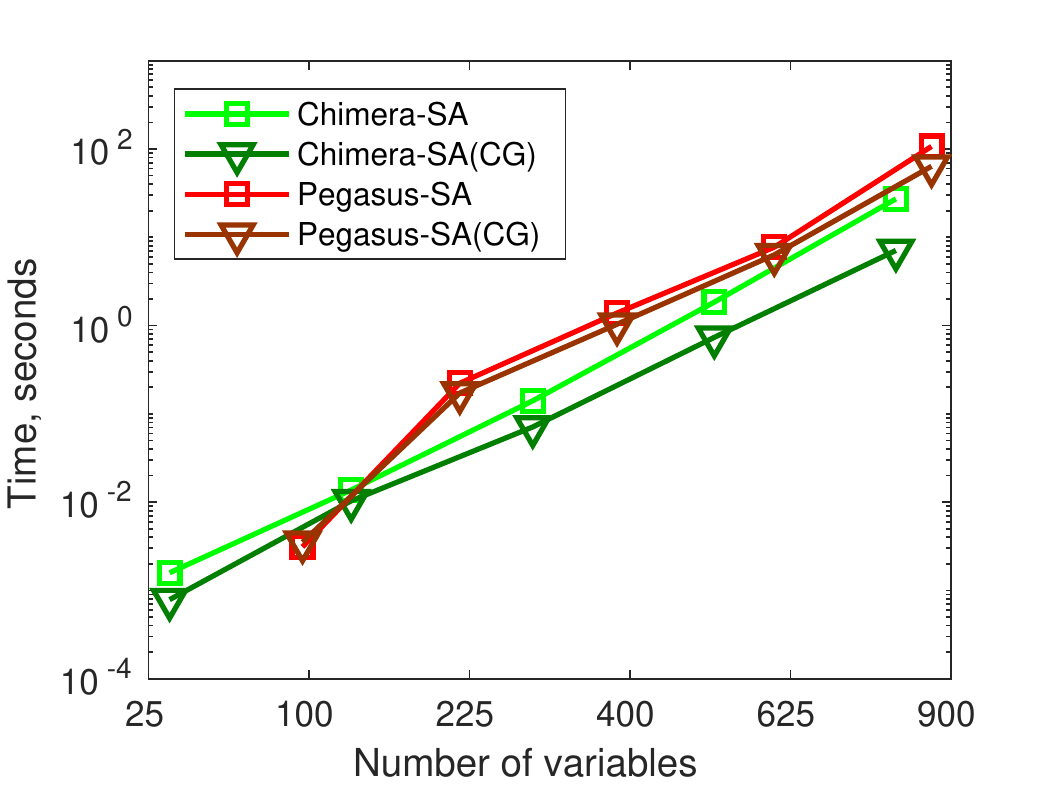}
\includegraphics[width=0.33\linewidth]{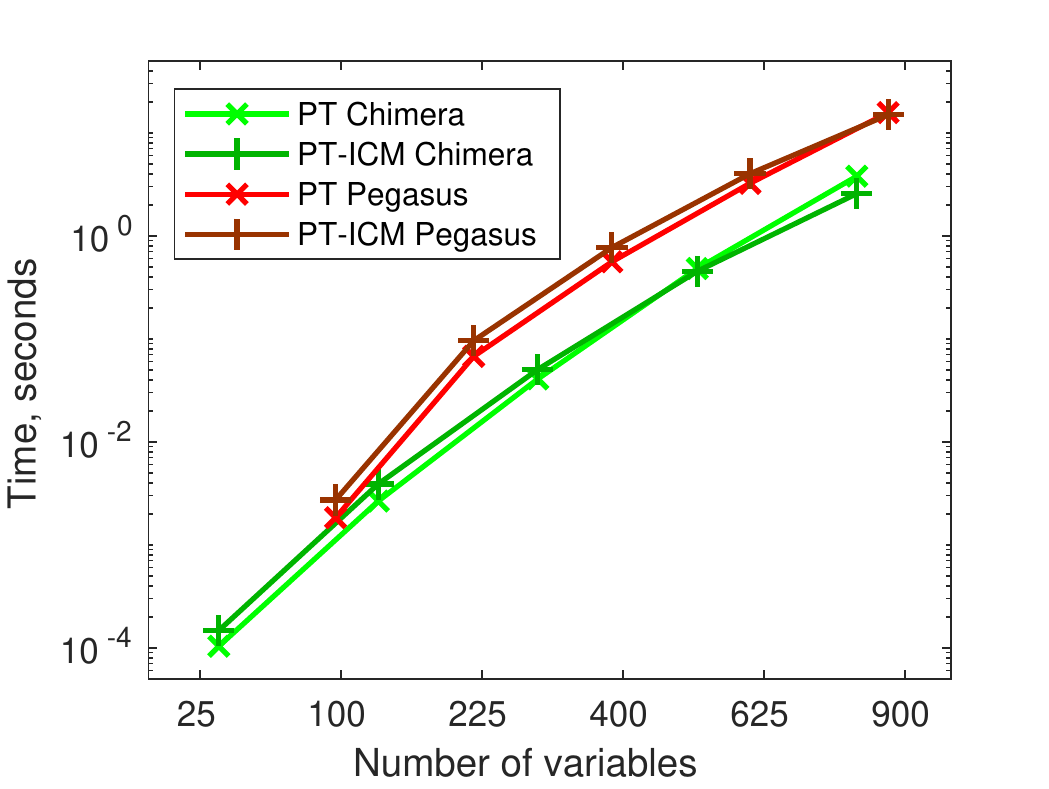}
\includegraphics[width=0.33\linewidth]{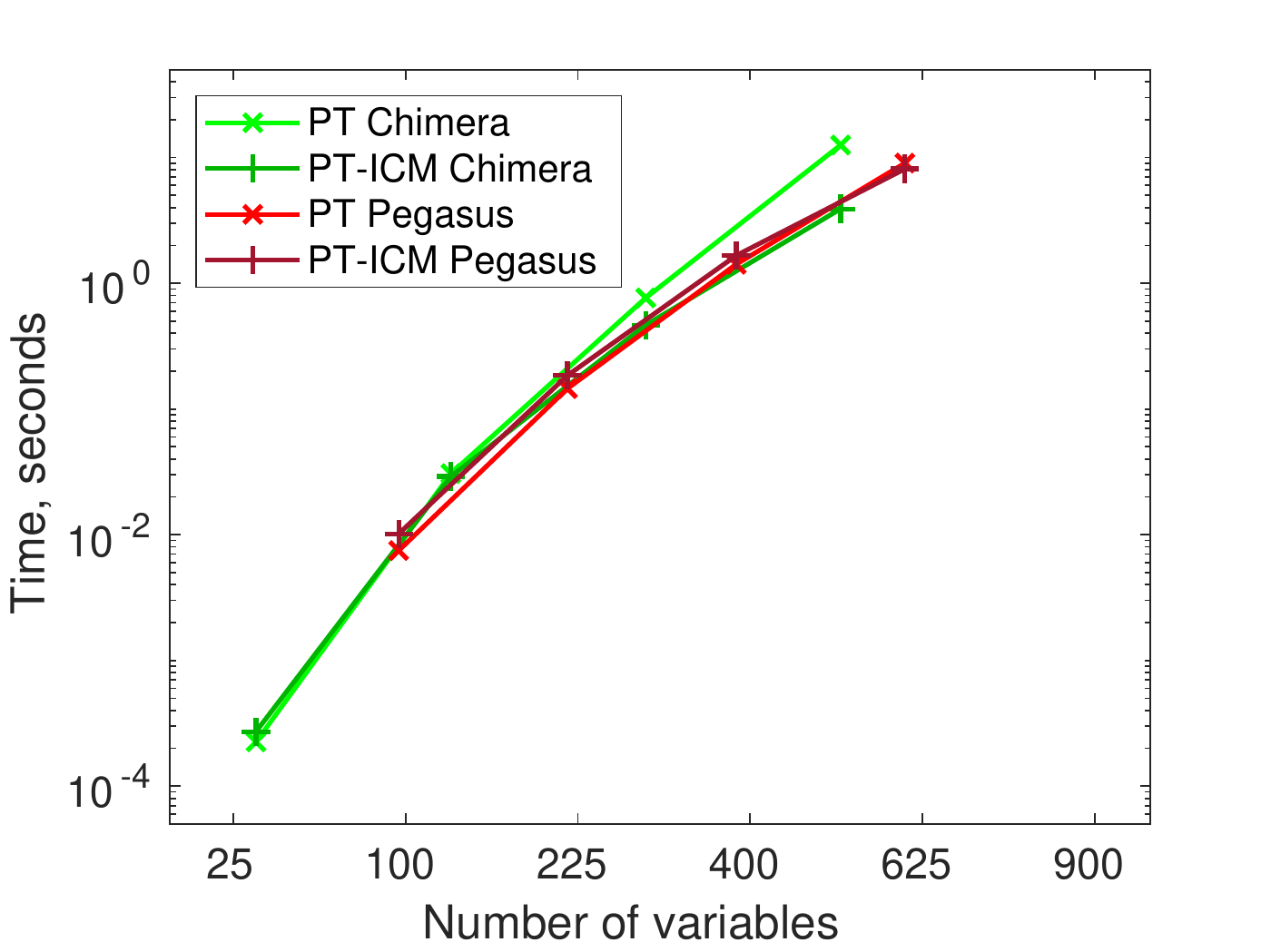}
\caption{\label{figureAlgs}
  Median time to solution for RAN1 and RANinf spin-glass problems running C++ implementations of several heuristics suitable for generic sparse graphs on single threaded 2.6 GHz CPUs. Bootstrapped error bars are small compared to marker size, and omitted.
  Left figure: simulated annealing on RAN1. For simple test cases with eight-qubit clusters, the impact of large-area local-search moves on TTS is more significant in Chimera. Typical TTS by rate-optimized simulated annealing heuristics is longer on Pegasus graphs. This difference becomes more pronounced in moving to RANinf instances.
  Center figure: parallel tempering on RAN1. Houdayer moves reduce typical TTS for Chimera graphs larger than 512 variables (C8), but are not useful in Pegasus graphs up to the sizes presented. Typical TTS by parallel tempering is longer on Pegasus graphs.
  Right figure: parallel tempering on RANinf. Houdayer moves enhance TTS for Chimera graphs at all scales, but have a small impact on Pegasus TTS. Typical TTS by vanilla parallel tempering is longer on Chimera graphs, but after exploiting the clustering present with ICM we see that there is no significant difference in TTS.}
\end{figure}

The eight-qubit sets are fixed and disjoint, covering all variables, and clusters consist of four vertical and four horizontal qubits located at minimum distance in space (in both Chimera and Pegasus). There is a hierarchy of large-area local-search moves of this kind known to perform very well in Chimera structured problems~\cite{1409.3934}. These moves are effective in Chimera because many of the interactions are captured within eight-qubit cell blocks. The area-move scale presented ``8'' is somewhat of a sweet spot both for Pegasus and Chimera, but fewer interactions are tidily packaged this way in Pegasus.
Our version of SA uses a geometric schedule, where annealing time (relative to restarts) has been optimized for each combination of algorithm, topology and system size.\footnote{Specifically $T=T_{max}(T_{min}/T_{max})^{i/n}$, with a fast mixing initial temperature $T_{max}=1/\sqrt{\sum_{ij} \langle J_{ij}^2 \rangle}$ and a final temperature with very low rates of excitation $T_{min}=-2/\log[0.99^{-1/N} -1]$, $N$ being the number of variables. The anneal is completed by a greedy descent from the final temperature, to ensure we do not miss local minima.} Figure \ref{figureAlgs} left side, shows the result for each method on RAN1 instances. Whereas eight-qubit cluster moves can accelerate solution search for Chimera, these have significantly less impact on Pegasus spin glasses, as expected.


Finally, we looked at the performance of parallel tempering, with two variations:

\begin{itemize}
    \item PT: without Houdayer moves~\cite{doi:10.1143/JPSJ.65.1604}
    \item PT-ICM: with Houdayer moves at all temperatures, also called iso-energetic cluster moves (ICM)~\cite{PhysRevLett.115.077201}
\end{itemize}

Single-spin updates used the Metropolis algorithm and the temperature ladder was tuned to good behaviour on the typical case for each system size and topology.\footnote{For the fixed interval $[T_{min},T_{max}]$ spacings were chosen to ensure mean replica exchange rate of $0.4$ across all instances.}

The advantages of Houdayer moves are intuitively limited by the percolation threshold and dimensionality, so we anticipate a stronger relative impact in Chimera. Figure \ref{figureAlgs}, center and right side, show results for the typical case: Pegasus RAN1 is more challenging than Chimera RAN1, whereas in RANinf, after using the ICM accelerator, typical behavior for Chimera and Pegasus is very similar. More importantly, the effectiveness of Houdayer moves is lower in Pegasus. Chimera RAN1 instances are in part easier than RANinf due to the degeneracy of the ground state, which we believe explains the discrepancy in TTS and lower sensitivity to ICM acceleration.

A good indicator of algorithmic hardness is the scaling of TTS: the larger the  instances for the problem classes and algorithms presented,  the slower to solve. Certain prefactors polynomial in the system size can be assumed in the TTS, but do not account for the major part of this scaling.
In these experiments we use a self-consistent test for the ground state. In Figure \ref{figureAlgs2} we marked the total run-time---as TTS approaches this value we can anticipate a significant bias (towards smaller TTS) owing to the ground state being misidentified, making it easier to "solve". We truncated the curves to mitigate for this bias; resampling indicates the bias is small in the results shown. Similarly, in figure \ref{figureAlgs} the ground state was determined self-consistently by comparing results over all algorithms, adjusting upwards the allowed run time for larger instances. At $N>1000$ variables, very long run times were required to mitigate for bias in harder instances, and curves are truncated accordingly.  

With accelerator tricks applied, Pegasus spin-glass problems require longer to solve than Chimera ones in the median, though this difference is negligible in the RANInf case for PT-ICM. This result should be interpreted with care: these are different problems and there is no obvious reason one ought to be invariably harder than the other, in the median and other quantiles.\footnote{We find evidence for some very hard RAN1 instances; for example, in the higher quartiles.} For example, lower connectivity in Chimera may make its energy barriers smaller and its local degeneracy higher, but it also makes it far less homogeneous in space.\footnote{There is less ``self-averaging'' of interactions, which can contribute to phenomena such as Griffiths singularities.} The more important points to acknowledge are that (1) we have flexibility in the Pegasus framework to model Chimera-like benchmarks, but not vice-versa, at least not without embedding; (2) Pegasus algorithm performance is less amenable to the standard accelerator tricks that have been so successful on Chimera.

\section{Conclusion}

The flexible architecture of D-Wave's next-generation of processors introduces qubits with a
higher degree of connectivity and a new type of coupler in a family of topologies, Pegasus.
This new topology has significant advantages over previous generations, including more
efficient embeddings for many useful classes of problems.

In this paper we describe, and provide a formulaic description for, the new topology. We describe methods
for translating known embeddings from Chimera to Pegasus and note some key advantages; for example:

\begin{itemize}
    \item Cliques and bicliques. Pegasus $P_M$ supports embedding cliques of up to size $12(M-1)$, with chains of length $M$ and $M+1$, and bicliques up to $K_{12M-20, 12M-20}$, with uniform chain length of $M-1$.
    \item Lattices. Pegasus supports a $(M-1) \times (M-1) \times 12$ cubic lattice in $P_M$ with uniform chainlength of $2$. We also show 2d lattice embeddings similar to those used for materials science research.
    \item Treewidth. Pegasus $P_M$ has a treewidth of between $12M-11$ to $12M-4$; for comparison, treewidth of a Chimera $C_M$ graph is $4M$.
    \item Odd couplers. We described how this new type of coupler, in addition to increasing connectivity, might be used in a simple error-correction scheme that increases the energy scale of the logical problem represented on the Pegasus topology.
\end{itemize}

We present results of our initial test of the topology in the following categories:

\begin{itemize}
    \item Heuristic embeddings. Our comparison of embeddings in $P_6$ (with 680 qubits and 4484 couplers) versus $C_{16}$ (with 2048 qubits and 6016 couplers) for a diverse set of problems shows Pegasus consistently achieving around a 50-60\% reduction in chainlength over Chimera.
    \item Native structured Ising models. Simple Pegasus benchmarks differ qualitatively from the corresponding Chimera ones, indicating that the topology may support larger energy barriers and thereby offer more opportunities for differentiation between classical and quantum dynamics. Techniques commonly in use to accelerate optimization (and sampling) in Chimera-structured problems seem less potent on Pegasus-structured problems.
\end{itemize}


\bibliographystyle{unsrt}
\bibliography{article}

\begin{thebibliography}{10}

\bibitem{cliquepaper}
Kelly Boothby, Andrew~D. King, and Aidan Roy.
\newblock Fast clique minor generation in chimera qubit connectivity graphs.
\newblock {\em Quantum Information Processing}, 15(1):495--508, 1 2016.

\bibitem{KlySulHum}
Christine Klymko, Blair~D. Sullivan, and Travis~S. Humble.
\newblock Adiabatic quantum programming: minor embedding with hard faults.
\newblock {\em Quantum Information Processing}, 13(3):709--729, 3 2014.

\bibitem{Robertson1986}
Neil Robertson and P.~D. Seymour.
\newblock {Graph minors. II. Algorithmic aspects of tree-width}.
\newblock {\em Journal of Algorithms}, 7(3):309--322, 9 1986.

\bibitem{Dechter99bucketelimination}
R~Dechter.
\newblock {Bucket elimination: A unifying framework for reasoning}.
\newblock {\em Artificial Intelligence}, 1--2:41--85, 1999.

\bibitem{Bodlaender2011}
Hans~L. Bodlaender and Arie~M.C.A. Koster.
\newblock {Treewidth computations II. Lower bounds}.
\newblock {\em Information and Computation}, 209(7):1103--1119, 7 2011.

\bibitem{Bodlaender2010}
Hans~L. Bodlaender and Arie~M.C.A. Koster.
\newblock {Treewidth computations I. Upper bounds}.
\newblock {\em Information and Computation}, 208(3):259--275, 3 2010.

\bibitem{Pudenz2014}
Kristen~L. Pudenz, Tameem Albash, and Daniel~A. Lidar.
\newblock {Error-corrected quantum annealing with hundreds of qubits}.
\newblock {\em Nature Communications}, 5:3243, feb 2014.

\bibitem{Vinci2016}
Walter Vinci, Tameem Albash, and Daniel~A Lidar.
\newblock {Nested quantum annealing correction}.
\newblock {\em npj Quantum Information}, 2(1):16017, nov 2016.

\bibitem{PhysRevE.96.022139}
R.~R.~P. Singh and A.~P. Young.
\newblock Critical and {G}riffiths-{M}c{C}oy singularities in quantum {I}sing
  spin glasses on $d$-dimensional hypercubic lattices: A series expansion
  study.
\newblock {\em Phys. Rev. E}, 96:022139, Aug 2017.

\bibitem{PhysRevX.4.021008}
Helmut~G. Katzgraber, Firas Hamze, and Ruben~S. Andrist.
\newblock Glassy chimeras could be blind to quantum speedup: Designing better
  benchmarks for quantum annealing machines.
\newblock {\em Phys. Rev. X}, 4:021008, Apr 2014.

\bibitem{PhysRevX.5.019901}
Martin Weigel, Helmut~G. Katzgraber, Jonathan Machta, Firas Hamze, and Ruben~S.
  Andrist.
\newblock Erratum: Glassy chimeras could be blind to quantum speedup: Designing
  better benchmarks for quantum annealing machines [phys. rev. x 4, 021008
  (2014)].
\newblock {\em Phys. Rev. X}, 5:019901, Jan 2015.

\bibitem{PhysRevLett.72.4141}
H~Rieger and A~P Young.
\newblock {Zero-temperature quantum phase transition of a two-dimensional Ising
  spin glass}.
\newblock {\em Phys. Rev. Lett.}, 72(26):4141--4144, jun 1994.

\bibitem{Kirkpatrick:OSA}
S~Kirkpatrick, C~D Gelatt, and M~P Vecchi.
\newblock {Optimization by Simulated Annealing}.
\newblock {\em Science}, 220(4598):671--680, 1983.

\bibitem{doi:10.1143/JPSJ.65.1604}
Koji Hukushima and Koji Nemoto.
\newblock {Exchange Monte Carlo Method and Application to Spin Glass
  Simulations}.
\newblock {\em Journal of the Physical Society of Japan}, 65(6):1604--1608,
  1996.

\bibitem{1409.3934}
Alex Selby.
\newblock Efficient subgraph-based sampling of {I}sing-type models with
  frustration, 2014.

\bibitem{PhysRevLett.115.077201}
Zheng Zhu, Andrew~J. Ochoa, and Helmut~G. Katzgraber.
\newblock Efficient cluster algorithm for spin glasses in any space dimension.
\newblock {\em Phys. Rev. Lett.}, 115:077201, Aug 2015.

\bibitem{Hamze2004}
Firas Hamze and Nando de~Freitas.
\newblock {From fields to trees}.
\newblock In {\em Proceedings of the 20th conference on Uncertainty in
  artificial intelligence}, pages 243--250. AUAI Press, jul 2004.

\bibitem{Gogate2004}
Vibhav Gogate and Rina Dechter.
\newblock {A Complete Anytime Algorithm for Treewidth}.
\newblock {\em Proceedings of the 20th conference on Uncertainty in artificial
  intelligence}, (1):201----208, 2004.

\end{thebibliography}

\appendix



\end{document}